# Analyse de l'hybridation dans les populations françaises de sangliers à l'aide de données de génotypage pangénomique

Nicolas MARY (1), Nathalie IANNUCCELLI (1), Geoffrey PETIT (2), Nathalie BONNET (1), Alain PINTON (1),
Vladimir GROSBOIS (2), Bertrand SERVIN (1), Juliette RIQUET (1), Alain DUCOS (1)

(1) GenPhySE, Université de Toulouse, INRAE, ENVT, 31326, Castanet Tolosan, France
(2) ASTRE, CIRAD, INRAE, 34398, Montpellier, France

**Analyse de l'hybridation dans les populations françaises de sangliers à l'aide de données de génotypage pangénomique**

Un suivi de la « pureté génétique » des populations françaises de sangliers est réalisé depuis les années 1980 en se basant sur l'existence d'une différence cytogénétique entre sangliers et porcs domestiques (nombres de chromosomes respectivement égaux à 36 et 38). Cette différence permet d'attribuer un statut « d'hybride » à tout sanglier contrôlé à 37 ou 38 chromosomes, sans toutefois pouvoir déterminer l'origine (récente ou ancienne) de l'hybridation, ni garantir la « pureté » d'un animal à 36 chromosomes. L'analyse des résultats de plus de 4600 contrôles réalisés au cours des 12 dernières années révèle un taux moyen « d'hybrides » de 15,8%, avec une forte variabilité selon les populations. Pour analyser plus finement l'hybridation en s'affranchissant des limites inhérentes à l'approche cytogénétique, 362 sangliers récemment prélevés dans différentes régions françaises ont été génotypés sur puce SNP 70K (GeneSeek GGP Porcine HD). Ces travaux montrent que pour 96,4% des sangliers analysés, incluant la plupart des animaux à 37 et 38 chromosomes, la proportion du génome d'origine « porc domestique » est comprise entre 0 et 18%. Ceci suggère que l'hybridation est un phénomène assez commun mais d'intensité modérée, et souvent ancienne. Des taux d'hybridation plus élevés ont néanmoins été observés dans certaines régions comme l'Ardèche, et des cas d'hybridations récentes avec des porcs domestiques ont été mis en évidence chez 3,6% des sangliers analysés, la plupart avec des porcs de type asiatique.

**Analysis of hybridization in French wild boar populations using genome-wide genotyping data**

The "genetic purity" of French wild boar populations has been monitored since the 1980s based on a cytogenetic difference between wild boars and domestic pigs (36 and 38 chromosomes, respectively). This difference makes it possible to identify any boar with 37 or 38 chromosomes as "hybrid", without however being able to determine the origin (recent or ancient) of the hybridization, nor guarantee the "purity" of an animal with 36 chromosomes. Analysis of results of more than 4,600 tests performed over the last 12 years reveals an average "hybrid" rate of 15.8%, with high variability between populations. To analyse hybridization in greater detail and overcome inherent limitations of the cytogenetic approach, 362 wild boars recently collected in different regions of France were genotyped on a 70K SNP (GeneSeek GGP Porcine HD) chip. This study showed that for 96.4% of the wild boars analysed, including most of those with 37 or 38 chromosomes, the percentage of the genome of "domestic pig" origin varied from 0 to 18%. This suggests that hybridization is a fairly common phenomenon but of moderate intensity, and often ancient. Nevertheless, higher rates of hybridization have been observed in some regions such as Ardèche, and several cases of recent hybridization with domestic pigs were found in 3.6% of the wild boars analysed, most of them with pigs of Asian origin.

## INTRODUCTION

Les porcs domestiques (*Sus domesticus,* Erxleben, 1777) sont issus de la domestication d'animaux sauvages (ou sangliers ; *Sus scrofa*, Linnaeus, 1758) [les noms repris ici sont conformes aux recommandations de l'International Commission on Zoological Nomenclature ; (Gentry *et al.*, 2004)]. La domestication du porc a été initiée indépendamment en Anatolie et dans la vallée du Mékong il y a environ 9 000 ans. En Europe, le processus de domestication a duré plusieurs millénaires et s'est accompagné de flux de gènes réguliers entre populations sauvages (sangliers européens différents de la population anatolienne impliquée dans le processus initial de domestication) et domestiques (Frantz *et al.*, 2015). Le développement progressif de l'élevage et la sélection des populations domestiques ont induit des différences génétiques et phénotypiques très importantes entre porcs domestiques et sangliers. Ces différences sont aussi importantes entre porcs domestiques européens et asiatiques, du fait (1) de critères de sélection très contrastés entre ces deux régions du monde, et (2) des différences existant entre les populations sauvages à partir desquelles la domestication s'est opérée (la divergence entre sangliers européens et asiatiques remonte à 1 million d'années (Groenen, 2016)).

Une différence génétique importante entre porcs domestiques et sangliers d'Europe de l'ouest concerne le nombre de chromosomes par cellule : 2n=38 et 2n=36, respectivement (Darre *et al.*, 1992; Aravena et Skewes, 2007). Le nombre réduit de chromosomes chez le sanglier s'explique assez simplement. Le caryotype du porc domestique contient deux paires de chromosomes acrocentriques : les paires 15 et 17. Chez le sanglier, les chromosomes de ces deux paires sont réunis en une seule paire de chromosomes submétacentiques (chacun des chromosomes de cette paire est constitué d'un chromosome 15 et d'un chromosome 17 de porc réunis par leurs centromères). En termes cytogénétiques, un sanglier à 2n=36 chromosomes est donc homozygote pour la translocation Robertsonienne 15/17 (il n'y a pas d'autre différence cytogénétique notable entre porc et sanglier). L'accouplement entre un porc domestique (2n=38) et un sanglier (2n=36) produit des individus hybrides à 37 chromosomes, qui héritent d'un chromosome de chacune des paires 15 et 17 de leur parent porc domestique, et d'un chromosome transloqué 15/17 de leur parent sanglier (on peut qualifier ces individus d'hétérozygotes pour la translocation). Ces animaux à 37 chromosomes peuvent être accouplés (1) à d'autres animaux à 37 chromosomes (donnant en espérance 25% d'individus à 36, 50% d'individus à 37 et 25% d'individus à 38 chromosomes), (2) à des animaux à 36 chromosomes (donnant en espérance 50% d'individus à 36 et 50% d'individus à 37 chromosomes), ou (3) à des individus à 38 chromosomes (donnant en espérance 50% d'individus à 37 et 50% d'individus à 38 chromosomes).

En France comme dans d'autres régions du monde, les populations de sangliers sont, depuis plusieurs décennies, en forte expansion. Cet accroissement démographique important s'explique en partie par des évolutions de l'environnement de vie des animaux (intensification des cultures, réduction du nombre de prédateurs ou augmentation globale des températures moyennes) (Root *et al.*, 2003; Touzot *et al.*, 2020), mais aussi par l'effet direct d'interventions humaines (création d'élevages et lâchers de sangliers pour les activités de chasse, agrainage, hybridation volontaire ou involontaire avec des porcs domestiques) (Khederzadeh *et al.*, 2019). Cet accroissement démographique pose de nombreux problèmes : dégâts aux cultures, augmentation des risques sanitaires pour les élevages et les populations des régions concernées, pollution génétique des populations sauvages, entre autres.

Pour prévenir le risque d'hybridation volontaire et évaluer en continu le statut des populations de sangliers, des contrôles de « pureté génétique » sont réalisés de longue date en France (populations naturelles et élevages) (Darre *et al.*, 1992). Les résultats des contrôles réalisés depuis 2008 sont présentés dans cet article. Cependant, la technique utilisée jusqu'à présent pour réaliser les contrôles (cytogénétique : comptage du nombre de chromosomes par cellule) présente des limites importantes. La première est qu'elle ne permet pas de garantir qu'un individu à 36 chromosomes n'est pas le produit d'hybridation(s) (un animal à 36 chromosomes peut être issu d'un accouplement entre deux individus à 37, par exemple). L'autre limite de l'approche cytogénétique est qu'elle ne permet pas, pour un individu à 37 chromosomes par exemple, de dater l'évènement (ou les évènements) d'hybridation qui en est (sont) à l'origine, ni de quantifier la proportion de génome d'origine « porc domestique » chez les individus qualifiés « d'hybrides » (individus à 37 ou 38 chromosomes) ou de « purs sangliers » (animaux à 36 chromosomes). Le génotypage de marqueurs moléculaires répartis sur l'ensemble du génome permet en théorie de surmonter ces limites (Goedbloed *et al.*, 2013). Dans cette étude, nous avons génotypé 362 sangliers de statuts chromosomiques différents et issus de différentes régions françaises à l'aide d'une puce porcine 70K SNP dans le but de caractériser plus finement l'introgression de porc domestique dans les populations françaises de sangliers.

## 1. MATERIEL ET METHODES

### 1.1. Analyse cytogénétique des populations françaises de sangliers

Des analyses de statut chromosomique sont demandées par des structures professionnelles diverses (élevages, fédérations de chasse, réserves naturelles, associations…). Les techniques mises en œuvre au niveau de notre laboratoire (plateforme de contrôle chromosomique de l'ENVT) pour réaliser ces analyses nécessitent la collecte de sang total sur tubes héparinés. Des cultures cellulaires sont initiées à partir de ces prélèvements. Après divers traitements cytogénétiques classiques (pour les détails, voir par exemple Ducos *et al.* (1998)), les préparations cellulaires sont étalées sur lames de verre et le nombre de chromosomes par cellule est compté pour au moins 10 cellules de chaque individu.

### 1.2. Echantillonnage des animaux génotypés dans le cadre de la présente étude

Sur un total de 362 sangliers génotypés, 283 proviennent d'échantillons de sang ayant fait l'objet d'une analyse cytogénétique à l'ENVT entre 2017 et 2019. Ces animaux ont été choisis afin d'être représentatifs de la diversité des structures ayant envoyé des échantillons à analyser au laboratoire. Parmi celles-ci, nous en avons identifié sept ayant demandé un nombre d'analyses relativement important au cours des 12 dernières années (> 35) et présentant des taux d'animaux à 36 chromosomes très élevés (> 97%). Ces structures (populations) ont été qualifiées de « pures » (ce sont celles pour lesquelles le risque d'hybridation semblait le plus faible *a priori* ; ce qualificatif a été repris sur la figure 4b). Sur les 283 échantillons, 56 en sont issus. Des échantillons supplémentaires (n=28) proviennent d'individus capturés au sein d'une réserve naturelle de l'ouest de la France (Deux-Sèvres) présentant depuis les années 2000 des taux élevés (> 40%) d'individus à 37 ou 38 chromosomes. Le reste des individus (complément à 283, soit 199) a été choisi de façon à disposer d'une bonne répartition des échantillons en France métropolitaine. Des échantillons complémentaires (79 biopsies d'oreille réalisées par des chasseurs entre 2013 et 2017) nous ont été transmis par les coordonnateurs d'un projet visant à étudier l'origine de cas de maladie de l'œdème dans des populations de sangliers sauvages en Ardèche (Petit *et al.*, 2020). L'ADN a pu être extrait de ces échantillons mais le contrôle du statut chromosomique des animaux n'a pas été réalisé (culture cellulaire impossible).

L'hybridation de sangliers sauvages avec des porcs domestiques de type « vietnamien » (animaux initialement « de compagnie » relâchés dans le milieu naturel par leurs propriétaires) a été suspectée à diverses reprises (Delibes et Delibes, 2013). Des échantillons (sang total et biopsie de peau) ont été prélevés sur deux cochons de type « vietnamien » et ont été intégrés à notre étude (les prélèvements ont été effectués à l'occasion de la castration chirurgicale de ces animaux réalisée à l'ENVT à la demande des propriétaires).

Aucun animal de notre étude n'a été élevé / tué / prélevé spécifiquement pour les besoins de notre projet, qui n'a donc pas requis d'autorisation explicite (en application de la directive 2010/63/EU).

### 1.3. Extraction d'ADN et génotypage

L'ADN a été extrait des échantillons de sang à l'aide du kit *Blood DNA Isolation* (Norgen), et à l'aide d'un protocole classique (lyse avec protéinase K et précipitation à l'éthanol) pour les biopsies d'oreille. Des tubes de 4 µl d'ADN dilué à 50 ng/µl ont été préparés pour les génotypages, réalisés sur la plateforme Génomique et Transcriptomique du CRCT (www.poletechno-crct.inserm.fr) à l'aide d'une puce GeneSeek Genomic Profiler (GGP) 70 K HD Porcine (Illumina Inc, USA) qui comprend 68516 SNP. Tous les sangliers avaient un call rate (% des SNP pour lesquels un génotype a pu être déterminé) > 0,90 avec une moyenne de 0,93.

### 1.4. Analyses préliminaires des données de génotypage

Les résultats des génotypages de sangliers et de porcs domestiques de type « vietnamien » réalisés dans le cadre de la présente étude, ainsi que ceux obtenus pour des porcs domestiques de races commerciales (Duroc, Landrace Français, Large White, Meishan et Piétrain) et locales (Basque, Bayeux, Gascon, Limousin, Porc Blanc de l'Ouest) dans le cadre de projets antérieurs de notre équipe (Muñoz *et al.*, 2019; Mercat *et al.*, 2020), ont été considérés dans notre étude. Ces porcs domestiques ont été génotypés avec les puces Porcine SNP60 (v1 ou v2 ; Illumina Inc, USA) ou GGP70K. Seuls les SNP communs aux trois puces, situés sur les autosomes et pour lesquels les taux de génotypes manquants étaient inférieurs à 10%, ont été conservés pour les analyses (40 240 SNP au total).

L'introgression étant susceptible d'induire un écart à l'équilibre de Hardy-Weinberg, les SNP n'étant pas à l'équilibre ont été conservés pour l'analyse. Le nombre d'animaux génotypés dans les races commerciales était très important (plusieurs milliers). De façon à limiter les temps de calculs et à disposer d'échantillons de taille relativement homogène, nous avons, dans un premier temps, sélectionné 100 individus dans chacune des races commerciales, en cherchant à maximiser la diversité génétique au sein des échantillons étudiés (utilisation de la fonction snpgdsIBD du package R SNPRelate (Zheng *et al.*, 2012)). Une analyse en composantes principales (ACP) a ensuite été réalisée à l'aide de la fonction snpgdsPCA du même package afin de visualiser les principaux groupes d'animaux (génétiquement proches intra-groupe, génétiquement distincts entre groupes).

### 1.5. Analyse de l'introgression porc-sanglier et de l'admixture

La méthode de « Bayesian clustering » implémentée dans le logiciel Admixture (Alexander *et al.*, 2009) a été utilisée pour quantifier la proportion des différentes origines ancestrales possibles dans le génome de chaque individu (admixture). Une analyse non supervisée a été réalisée en utilisant les réglages par défaut pour des valeurs de K variant de 2 à 25. La valeur optimale de K (définissant le nombre de populations ancestrales distinctes, ou « clusters ») a été estimée à l'aide d'une procédure de validation croisée. La valeur retenue (K=11) correspond à celle pour laquelle la courbe de validation croisée atteint un plateau. Le package R pophelper (Francis, 2017) a été utilisé pour visualiser les résultats (Figure 3). La précision (erreurs standards et intervalles de confiance au risque de 95%) des estimations d'origine ancestrale (proportion du génome de chaque individu ayant pour origine chacun des 11 clusters précédemment définis) a été calculée à l'aide d'une procédure d'échantillonnage aléatoire (bootstrap ; 200 réplications). Les origines ancestrales ont été comparées entre sous-populations de sangliers à l'aide des tests de Kruskall-Wallis et Wilcoxon.

## 2. RESULTATS

### 2.1. Analyse cytogénétique des populations françaises de sangliers

Le nombre d'analyses cytogénétiques de sangliers réalisées annuellement à l'ENVT entre 2008 et 2019 a varié entre 160 et 576, avec une moyenne de 389. La diminution du nombre d'analyses est assez nette depuis le début des années 2010 (Figure 1). Le taux « d'hybrides » (sangliers hybrides ou issus d'hybrides à 37 ou 38 chromosomes) calculé à partir des résultats obtenus pour les 4 666 individus analysés dans notre laboratoire au cours de cette période est de 15,8% (13,4% à 2n=37 et 2,4% à 2n=38). Ce taux varie assez fortement d'une année à l'autre (Figure 1), en raison des fluctuations d'échantillonnage (ce ne sont pas nécessairement les mêmes structures professionnelles qui sollicitent des analyses d'une année sur l'autre, et les taux d'hybrides varient fortement d'une structure à l'autre : de 0% à 40% pour les structures ayant effectué au moins 25 analyses entre 2008 et 2019). L'évolution au cours du temps de la moyenne des moyennes calculées pour les différentes structures sollicitant des analyses révèle néanmoins une augmentation du taux « d'hybrides » au cours de la période (Figure 1).

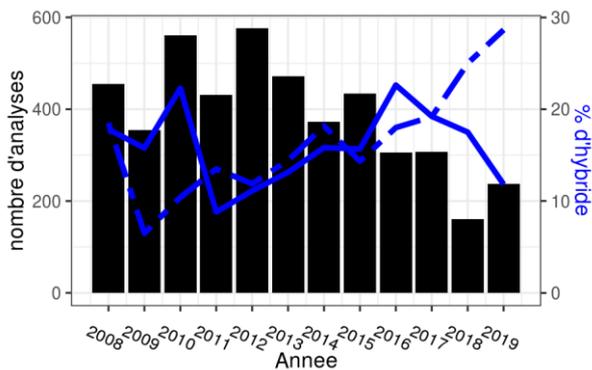

**Figure 1** – Evolution du nombre d'analyses cytogénétiques de sangliers réalisées entre 2008 et 2019, et de la fréquence d'animaux à 2n=37 ou 38 chromosomes (fréquence brute en trait plein, moyenne des moyennes par structure ayant envoyé des échantillons en trait pointillé).

### 2.2. Structure des populations de suidés et hybridation

Les deux premiers axes de l'ACP expliquent 10,5% et 5,5% de la variance totale, respectivement (Figure 2). Le gain de variance expliquée devient très faible à partir de la composante principale No 11, ce qui correspond au nombre de races analysées (en considérant conjointement les Meishan et les porcs de type « vietnamien ») et au nombre de clusters dans l'analyse d'admixture (voir le chapitre « matériel et méthodes »). La première composante principale permet de séparer trois groupes d'animaux. Le premier (noté A sur la figure 2) est constitué des individus des races asiatiques, le second regroupe les animaux des races européennes (B et B'), tandis que le dernier groupe est constitué de sangliers uniquement (C). La deuxième composante principale permet elle aussi de discriminer les sangliers des races asiatiques et européennes, mais également de distinguer la race Duroc qui forme un groupe distinct des autres races européennes (confirmation de résultats antérieurs ; voir par exemple (Lee et al., 2020)). Les sangliers occupant sur l'ACP une position intermédiaire entre les principaux groupes (notés GISA-XXX sur la figure 2) sont des « hybrides probables » (voir ci-après).

L'analyse de l'admixture pour les 362 sangliers de notre échantillon a montré que, pour 13 d'entre eux (soit 3,6%), issus de six structures détentrices différentes, les origines « porc domestique » étaient supérieures à 25%. Ces animaux ont été qualifiés « d'hybrides probables » (animaux issus d'évènement(s) d'hybridation pouvant être assez récent(s)), et la composition ancestrale de leurs génomes est représentée sur la figure 3. La majorité de ces « hybrides probables » (9/13) ont une part importante de leur génome (>35%) d'origine « asiatique » (en vert sur la figure 3). Trois autres (552, 553 et 554) ont une composition génomique assez comparable (présence d'une mosaïque d'origines ancestrales européennes, quasi-absence d'origine ancestrale asiatique), suggérant des hybridations pouvant être anciennes avec plusieurs races d'origine européenne. Un hybride (544) a 53% (± 5%) de son génome d'origine Gasconne (cas de figure qui ne se retrouve chez aucun autre individu de cette catégorie). Enfin, et de façon assez surprenante, l'un des « hybrides probables » (N° 516, contrôlé à 36 chromosomes lors de l'analyse cytogénétique) ne présente que 2,2% (± 1,8%) de son génome d'origine « sanglier », suggérant qu'il est le produit de nombreux croisements en retour avec des races de porcs domestiques.

La plupart de ces hybrides (l'animal 544 faisant exception) occupent bien sur l'ACP une position intermédiaire entre les principaux groupes raciaux (cf plus haut), ce qui est cohérent avec la composition « mosaïque » de leurs génomes (Figures 2 et 3).

Pour les 349 autres sangliers étudiés dans le cadre de ce projet, la proportion du génome d'origine « sanglier » varie entre 82 et 100% (moyenne et médiane égales à 94%). Comme le montre la figure 4a, cette proportion est supérieure ($P < 0{,}05$) chez les individus à 36 chromosomes (médiane de 96%) à celle que l'on trouve chez les individus à 37 ou 38 chromosomes (médianes de 93 et 90% respectivement).

Pour les 56 individus issus des sept structures les plus « pures » de notre échantillon (voir le chapitre « Matériel et méthodes »), la proportion du génome d'origine « sanglier » est de 98% (valeur médiane). Cette proportion est aussi très élevée (valeur médiane de 99%) dans l'échantillon de sangliers des Deux-Sèvres présentant pourtant un taux d'animaux à 37 et 38 chromosomes assez élevé (figure 4b). Elle est significativement plus faible (médiane de 92%) pour les animaux sauvages prélevés en Ardèche (Figure 4b).

L'origine ancestrale des génomes des deux animaux de type « vietnamien » est assez proche de celle des animaux de race Meishan, confirmant leur origine asiatique (Figures 2 et 3).

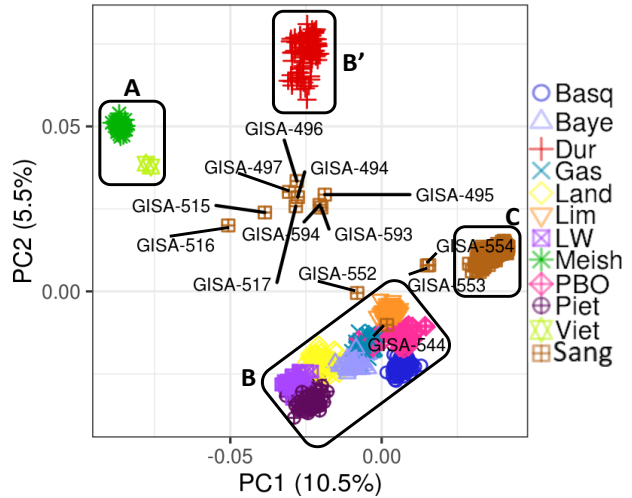

**Figure 2** – Résultats de l'Analyse en Composantes Principales permettant de distinguer les différents types génétiques (ACP).

### 3. DISCUSSION

La pureté génétique des populations sauvages et d'élevage de sangliers en France suscite l'intérêt des chercheurs, des gestionnaires de faune sauvage et des pouvoirs publics depuis de nombreuses années. Les études rétrospectives à grande échelle réalisées jusqu'à présent dans notre pays se basaient sur des analyses cytogénétiques uniquement (Darre et al., 1992; Ducos et al., 2008). Pour la première fois, cette question est abordée à l'aide de données de génotypage moléculaire à haute densité. Les seuls travaux comparables réalisés en France dans le passé se basaient sur le génotypage d'un panel de 20 marqueurs SNP, ne permettant pas la même profondeur d'analyse (Beugin et al., 2017). Les résultats montrent que, globalement, la proportion des génomes de sangliers d'origine « porc domestique » et la proportion d'animaux « hybrides probables » au sein de notre échantillon de sangliers sont assez faibles (de l'ordre de 6% et 4%, respectivement). Ces résultats sont cohérents avec ceux obtenus en Europe du Nord (Allemagne, Belgique et Pays-Bas) par Goedbloed *et al.* (2013) à l'aide de méthodes comparables (pourcentage d'hybrides égal à 3,9%). Ils sont par contre en décalage assez important avec les résultats issus des analyses cytogénétiques (15,8 % de sangliers « hybrides » à 2n=37 ou 38). Cette différence peut s'expliquer par le fait que des animaux à 37 chromosomes (voire à 38) peuvent être issus d'évènements d'hybridation (avec des porcs domestiques) relativement anciens, suivis de croisement en retour vers le sanglier assez nombreux, contribuant à réduire (diluer), génération après génération, le pourcentage de gènes d'origine « porc domestique » chez les descendants (tout en conservant à chaque génération, comme évoqué en introduction, une probabilité non nulle de procréer des descendants à 37 chromosomes).

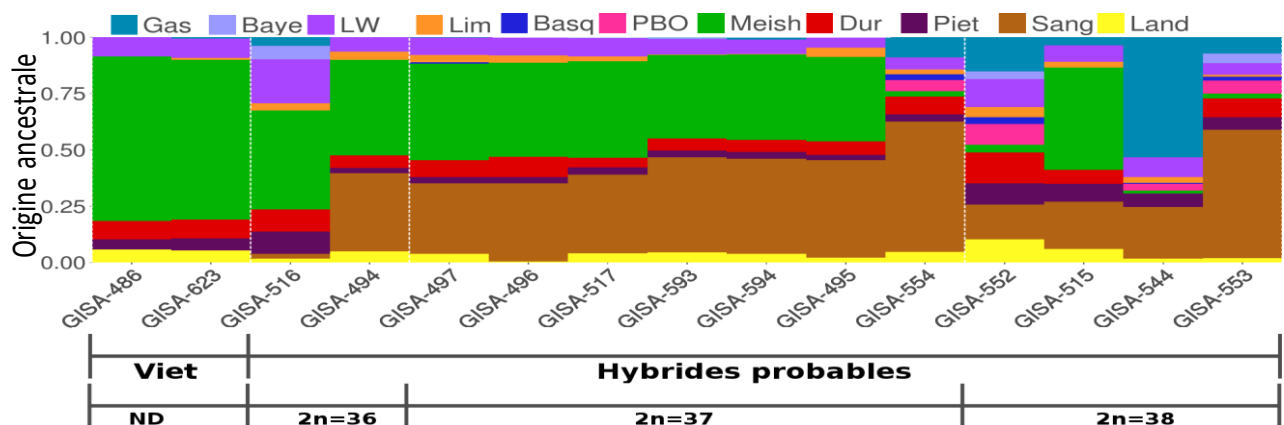

**Figure 3** – Compositions génomiques des sangliers ayant au moins 25% de leur génome d'origine « porc domestique » (+ porcs domestiques de type « vietnamien » à gauche de la figure). Les différentes couleurs représentent les origines ancestrales dominantes des différentes races ou groupes génétiques considérés (K=11).

Ce décalage est encore plus important pour les animaux issus de la réserve naturelle des Deux Sèvres. Les premières études de statut chromosomique réalisées en 1989-1990 dans cette population n'avaient détecté aucun sanglier hybride à 2n=37 ou 38. Sur la période récente (2008-2019), le pourcentage d'hybrides (2n=37 ou 38) est assez élevé (40%), alors que la proportion de génome d'origine « porc domestique » chez ces animaux est très faible (1%). Cette population de sangliers des Deux-Sèvres a la particularité de provenir d'une forêt domaniale protégée des activités sylvicoles et classée réserve nationale de chasse et de faune sauvage depuis 1973. Les tempêtes « Lothar » et « Martin » qui ont touché la France en décembre 1999 pourraient expliquer l'augmentation rapide du nombre d'individus hybrides (2n=37) au début des années 2000. En effet, la destruction des clôtures induite par ces tempêtes dévastatrices a pu faciliter les interactions avec des porcs d'élevage et/ou de compagnie. Ces hybridations accidentelles auraient été suivies pendant les 20 années suivantes de plusieurs générations de croisement en retour contribuant à diluer rapidement et fortement la proportion de gènes d'origine « porc domestique » dans cette population.

L'analyse des animaux issus de la population sauvage d'Ardèche montre une proportion de génome d'origine « porc domestique » sensiblement plus élevée (7% en moyenne chez ces animaux). Ces résultats pourraient être la traduction d'hybridations avec des porcs domestiques plus fréquentes dans cette région, pouvant éventuellement expliquer, en partie, l'accroissement démographique important de ces populations de sangliers et l'émergence de pathologies telles que la maladie de l'œdème (maladie fréquente chez le porc, mais jamais décrite dans les populations sauvages de sangliers avant leur identification en Ardèche en 2013 ; Petit *et al.*, 2020).

Globalement, aucun « hybride probable » (animaux ayant plus de 25% de leur génome d'origine ancestrale « porc domestique ») n'a été détecté chez les 110 individus provenant des deux populations de sangliers évoluant en liberté (espaces naturels en Ardèche et dans les Deux-Sèvres). Le taux d'hybridation est légèrement plus important dans les structures de type « élevage » ou « parc de chasse », et l'utilisation de reproducteurs de type génétique « asiatique » semble être la principale source d'hybridation récente (en raison peut être de la plus grande ressemblance phénotypique avec le sanglier). Le recours à l'hybridation avec des porcs domestiques est susceptible d'avoir des effets positifs dans ce type de structure (réduction de la consanguinité, valorisation d'effets d'hétérosis et d'effets additifs des gènes importants pour certains caractères comme la prolificité) (Iversen *et al.*, 2019; Frankham, 1995). Cette pratique peut donc être assez tentante pour les gestionnaires de ces structures. Même si nos résultats suggèrent qu'elle est probablement assez rare, la poursuite d'un contrôle rigoureux semble nécessaire et importante.

Le cas de l'individu 516 est très particulier. Cet animal présente un caryotype de sanglier (2n=36) mais un génome de porc (mosaïque d'origines ancestrales « porc domestique », dont 44% (±5%) d'origine « asiatique »). Le sexe et le nombre de chromosomes déterminés à partir des données de génotypage (origine « sanglier » des régions centromériques des chromosomes 15 et 17 suggérant la fusion des chromosomes des deux paires chez cet individu) sont cohérents avec les résultats cytogénétiques. Ceci ne milite donc pas en faveur de l'hypothèse d'une erreur d'identification. Cet animal a été déclaré comme sanglier dans son élevage d'origine. Le vétérinaire ayant réalisé le prélèvement n'a pas déclaré de particularité phénotypique chez cet animal (de nature à jeter un doute quant à son origine « sanglier »), ce qui est très surprenant compte-tenu de la composition de son génome. Cet animal ayant été vendu pour la chasse par son propriétaire, aucune expertise complémentaire n'est malheureusement envisageable.

## 4. CONCLUSION

Le décalage important entre les résultats des analyses cytogénétiques (comptage chromosomique) et de génotypage moléculaire à haute densité (analyse de l'admixture) constaté dans notre étude, et les informations complémentaires qu'apportent le génotypage en termes d'origine ancestrale des génomes et d'ancienneté des évènements d'hybridation, par exemple, militent en faveur d'une évolution des méthodes de contrôle de la pureté génétique des populations de sangliers en France. Les outils existent et sont couramment utilisés dans les programmes de sélection des races porcines commerciales. Leur déploiement pour l'analyse génomique de sangliers faciliterait en outre les opérations de prélèvements (possibilité de réaliser des analyses à partir de poils, de biopsies de peau, éventuellement réalisées sur des animaux morts). Reste maintenant à organiser la transition technologique d'un point de vue pratique et logistique (en identifiant le(s) laboratoire(s) susceptibles de proposer ce type de service, en définissant un modèle économique pertinent…). Dans cette attente, le contrôle cytogénétique devrait être poursuivi. Nos résultats montrent en effet que la proportion du génome d'origine « porc domestique » est en moyenne plus faible chez les animaux issus de structures n'élevant que des animaux à 2n=36. Etre en mesure de maintenir un tel statut dans la durée est donc une assez bonne garantie (mais pas une garantie absolue) de « pureté génétique » des animaux produits.

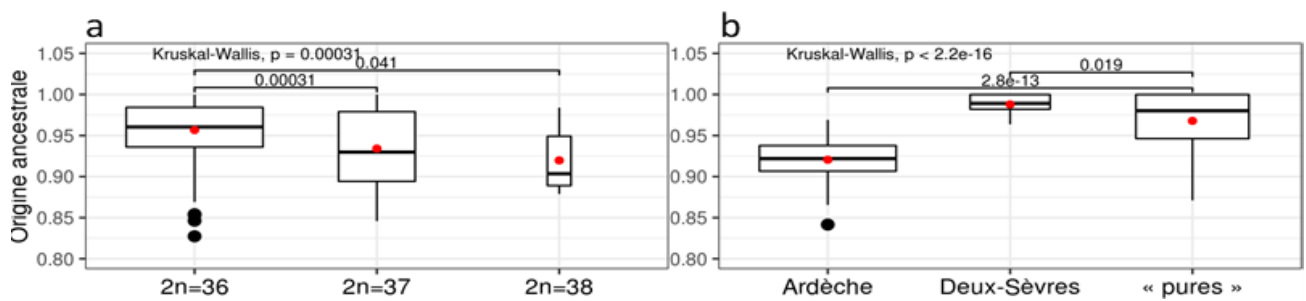

**Figure 4** – Comparaison de la proportion de génome d'origine ancestrale « sanglier » en fonction a) du statut chromosomique et b) de l'origine géographique et du type de structure d'origine des sangliers génotypés. Les points rouges indiquent les moyennes pour chaque catégorie (les 13 « hybrides probables » n'ont pas été considérés dans cette analyse).

## REMERCIEMENTS

## REFERENCES BIBLIOGRAPHIQUES